\def\Tr{{\rm Tr}}
\def\noblackbox{\overfullrule=0pt}
\noblackbox
\def\half{{1\over 2}}
\documentclass{ws-procs9x6}

\begin{document}

\title{Yangian Symmetry in D=4 Superconformal Yang-Mills Theory\footnote
{
\uppercase{QTS}3 \uppercase{c}onference \uppercase{p}roceedings, 
\uppercase{U}niversity of \uppercase{C}incinnati,
\uppercase{S}eptember 2003.}}

\author{Louise Dolan\footnote{\uppercase{W}ork partially
supported by \uppercase{DOE} grant \uppercase{DE-FG02-03ER-41262}.}}

\address{Department of Physics\\
University of North Carolina, Chapel Hill, NC 27599-3255\\}

\author{Chiara R. Nappi\footnote{\uppercase{T}his work is supported
by \uppercase{NSF}\uppercase{G}rants
\uppercase{PHY-0140311} and \uppercase{PHY-0243680}.}}

\address{Department of Physics, Jadwin Hall\\
Princeton University, Princeton, NJ 08540\\}

\author{Edward Witten\footnote{\uppercase{T}his work is supported
by \uppercase{NSF} \uppercase{G}rant \uppercase{PHY-0070928}}}

\address{School of Natural Sciences, Institute for Advanced Study\\
Olden Lane, Princeton, NJ 08540\\}


\maketitle

\abstracts{We will discuss an integrable structure for weakly
coupled superconformal Yang-Mills theories, describe certain
equivalences for the Yangian algebra, and fill a technical gap in
our previous study of this subject.}
\section{Symmetry Generators and Anomalous Dimensions}
In [\refcite{Bena}] it was shown that the classical Green-Schwarz
superstring action for $AdS^5\times S^5$ possesses a hierarchy of
non-local symmetries of the type that exist in integrable field
theories [\refcite{DB}, \refcite{NM}]. This is due to the fact
that the Green-Schwarz superstring in $AdS^5\times S^5$ can be
interpreted as a coset theory where the fields take values in the
coset superspace
\begin{equation}
{PSU(2,2|4)\over {SO(4,1)\times SO(5)}}\,.
\label{coset}
\end{equation}
This coset theory (even though the target is not a symmetric space
[\refcite{Wadia}]) admits non-local currents which give rise to
charges satisfying a Yangian algebra. A Yangian algebra $Y(G)$ is
an associative Hopf algebra [\refcite{{DRone},{DRtwo},{DB},{HH}}]
generated by the elements $J^A$ and $Q^B$ with
\begin{eqnarray}
&& [J^A, J^B]= f^{AB}_{C}J^C\,,\qquad [J^A, Q^B]= f^{AB}_C Q^C\,,
\label{yang}
\end{eqnarray}
and the Serre relations
\begin{eqnarray}
&&[Q^A, [Q^B, J^C]] + [Q^B, [Q^C, J^A]] + [Q^C, [Q^A, J^B]]\nonumber\\
&&\hskip5pt ={1\over 24}f^{ADK}f^{BEL}f^{CFM}f_{KLM}
\{J_D, J_E, J_F\}\,,
\label{yangs}\\
&&[[Q^A, Q^B], [J^C, Q^D]] + [[Q^C, Q^D], [J^A, Q^B]]\,, \nonumber\\
&& = {1\over 24} ( f^{AGL} f^{BEM} f^{KFN} f_{LMN} f^{CD}_K\nonumber\\
&&\hskip30pt
+ f^{CGL} f^{DEM} f^{KFN} f_{LMN} f^{AB}_K )  \{J_G, J_E, J_F\}\,,
\label{yangf}
\end{eqnarray}
for $J^A$ taking values in the Lie algebra of   an arbitrary
semi-simple Lie group $G$. (Lie algebra indices $A,B,C$ are raised
and lowered with an invariant, nondegenerate metric tensor
$g_{AB}$ or $g^{AB}$.) The symbol $\{A,B,C\}$ denotes the
symmetrized product of three operators $A,B,$ and $C$. Under
repeated commutators, the $Q^A$ generate an infinite-dimensional
symmetry algebra that has been called the Yangian.   The Yangian
has a basis ${\mathcal{J}}_n^A$ where ${\mathcal{J}}_0^A=J^A$,
${\mathcal{J}}_1^A=Q^A$, and ${\mathcal{J}}_n^A$ is an $n$-local
operator that arises in the $(n-1)$-form commutator of $Q$'s.
Since we will work in this paper mainly with the generators $J^A$
and $Q^A$, we have given them those special names. The Yangian
relations as written above are redundant in the following sense.
For $SU(2)$ the relation (\ref{yangs}) is trivial. For other cases
such as $SU(N)$ with $N\ge\nobreak3$, the relation (\ref{yangs})
implies the following one (\ref{yangf}).

In the superstring on $AdS^5\times S^5$, the $J^A$ will be
generators of $G = PSU(2,2|4)$, the brackets will generalize to
denote either commutators or anticommutators, and $f^C_{AB}$
become the structure constants of $PSU(2,2|4)$. The $Q^A$ are the
new non-local charges whose existence was found in
[\refcite{Bena}]. If the AdS/CFT correspondence is correct, then
on the CFT side we must have the same Yangian symmetry. The
question arises of what could be the $Q^A$ charges in the super
Yang-Mills side. We will answer this question in the extreme weak
coupling limit, that is, the opposite limit from that which is
considered in [\refcite{Bena}].  In order to justify our guess for
$Q^A$, we will review how non-local symmetries arise in
two-dimensional sigma models.

Let us  consider a model with a group $G$ of symmetries; the Lie
algebra of $G$ has generators $T_A$ obeying
$[T_A,T_B]=f_{AB}^CT_C$.  The action of $G$ is generated by a
current $j^{\mu\,A}$ that is conserved, $\partial_\mu j^{\mu\, A}
=0$.  Non-local charges arise if, in addition, the Lie algebra
valued current $j_\mu = \sum_A j_\mu^ A T_A$ can be interpreted as
a flat connection,
\begin{equation}
\label{flatco}
{\partial_\mu j_\nu-\partial_\nu j_\mu + [j_\mu ,
j_\nu]=0.}
\end{equation}
(Indices of $j_\mu$ are raised and lowered using the
Lorentz metric in two dimensions.) The conservation of $j_\mu$
leads in the usual fashion to the existence of conserved charges
that generate the action of $G$:
\begin{equation}
\label{hobbo}
{J^A=\int_{-\infty}^\infty dx \,j^{0\,A}(x,t).}
\end{equation}
In addition, a short computation using (\ref{flatco}) reveals that
\begin{equation}
\label{jobbo}
{Q^A = f^A_{BC} \int_{-\infty}^\infty dx\,\int_x^\infty
dy\, j^{0\,B}(x,t) \, j^{0\,C}(y,t) - 2 \int_{-\infty}^\infty dx
\, j_1^A(x,t) }
\end{equation}
is also conserved. The charges $J^A$ and $Q^A$ generate a Yangian
algebra [\refcite{{DB},{NM}}], even though the infinitesimal
transformations generated by them generate half of a Kac-Moody
algebra [\refcite{D}].


There are also discrete spin systems, that is systems in which the
dynamical variables live on a one-dimensional lattice rather than
on the real line, that similarly have Yangian symmetry.  The
lattice definition of $J^A$ is clear.  We assume that the spins at
each site $i$ have $G$ symmetry, and transform in some
representation ${\mathcal R}$.   We let $J^A_i$ be the symmetry operators at
the $i^{th}$ site. The total charge generator for the whole system
is then
\begin{equation}
\label{uncu}
{J^A=\sum_i J^A_i.}
\end{equation}
What about $Q^A$?  For general $G$ and ${\mathcal R}$, there is no
satisfactory definition of $Q^A$.  However, for $G=SU(N)$, a $Q^A$
can be defined for any ${\mathcal R}$.  For certain representations, the
requisite formula for $Q^A$ is particularly simple.  One just
takes the obvious discretization of the  bilocal part of
(\ref{jobbo}):
\begin{equation}\label{nobbo}
{Q^A=f^A_{BC}\sum_{i<j}J^B_iJ^C_j.}
\end{equation}
This is  the right formula in many of the most commonly studied
lattice integrable systems.

In (\ref{nobbo}), 
one has made no attempt to discretize the second term
in (\ref{jobbo}).  In fact, for many choices of ${\mathcal R}$, discretizing
that second term is impossible for elementary reasons.  One would
expect a discretization of $\int dx \,j_1^A$ to be of the form
$\sum_i j_i^A$ where $j_i^A$ acts on the $i^{th}$ site and
transforms in the adjoint representation. If ${\mathcal R}$ is such that the
adjoint representation of $G$ appears only once in the
decomposition of ${\mathcal R}\otimes\overline {\mathcal R}$, 
then $j_i^A$ would have to
be a multiple of $J_i^A$.  This is so, for example, if ${\mathcal R}$ is the
fundamental representation of $SU(N)$ or $SU(N|M)$ (and more
generally if ${\mathcal R}$ is the representation of $k^{th}$ rank
antisymmetric tensors, for any $k$). But taking $j_i^A$ to be a
multiple of $J_i^A$ just adds to $Q^A$ a multiple of $J^A$; this
is an outer automorphism of the Yangian algebra, and so does not
help (or hurt) in obeying the Serre relations.  We will argue that
(\ref{nobbo}) is the correct formula for the $Q^A$ in the weak
coupling limit of Yang-Mills theory.  This will fill a technical
gap in our previous paper [\refcite{DNW}], where we showed that
$Q^A$ as defined by this formula commutes with the one-loop
anomalous dimension operator, leading to an infinity of
conservation laws, but we did not show that it obeys the Serre
relations, which is needed to ensure that the resulting integrable structure is the
standard Yangian algebra.

Another justification for the proposal
(\ref{nobbo}) is that it has an analog
in gauge field theory at $g^2N=0$.
In order to make contact with conventional Noether current
symmetry analysis, we give the expression for the non-local charge
(\ref{nobbo}) in terms of the elementary fields of the super Yang-Mills
Lagrangian
\begin{equation}
\label{labo}
{\mathcal L} = {1\over {g_{YM}^2}}\, \Tr \left(
{1\over 2}F_{\mu\nu} F^{\mu\nu} + D_\mu\phi^I D^\mu\phi^I - {1\over 2}
[\phi^I, \phi^J ]  [\phi^I, \phi^J ] + {\rm fermions}\right)\,.
\end{equation}
For simplicity, we will only consider $A\in so(2,4)$.
In the classical  theory, the symmetry
currents for the conformal
group are given in terms of the improved energy-momentum tensor by
\begin{equation}
\label{noeth}
j^{A\mu} (x) = \kappa^A_\nu \theta^{\mu\nu}(x)\,,
\end{equation}
where $\kappa^A_\mu$ are the conformal Killing vectors, and
\begin{eqnarray}
\theta^{\mu\nu}&&=  2 \Tr F^{\mu\rho} F^\nu_{\hskip2pt\rho}
+ 2 \Tr D^\mu \phi^I D^\nu \phi^I - g^{\mu\nu} {\mathcal L} \,
- {1\over 3} \Tr (D^\mu D^\nu - g^{\mu\nu} D_\rho D^\rho ) \phi^I \phi^I
\nonumber\\
&&\,\,+\,{\rm fermions}\,.
\label{niem}
\end{eqnarray}
The currents (\ref{noeth}) are conserved at any
$g^2N$ using the classical interacting equations of motion.
If we set $g^2N=0$, we note that the {\it untraced} matrix
\begin{eqnarray}
&&  (\theta^{\mu\nu})_i^{\hskip2pt j} =
F^{\mu\rho}
F^\nu_{\hskip2pt \rho}
+ F^{\nu\rho} F^\mu_{\hskip2pt \rho}
+ \partial^\mu \phi^I \partial^\nu \phi^I \cr &&
\hskip40pt + \partial^\nu \phi^I \partial^\mu \phi^I
- g^{\mu\nu} ( {1\over 2} F_{\rho\sigma} F^{\rho\sigma}
+ \partial_\mu \phi^I \partial^\mu \phi^I )\cr &&
\hskip35pt - {1\over 3} (\partial^\mu \partial^\nu - g^{\mu\nu}
\partial_\rho \partial^\rho ) \phi^I \phi^I
+ {\rm fermions}
\label{nim}
\end{eqnarray}
is also conserved, as is
$\kappa^A_\nu (\theta^{\mu\nu})_i^{\hskip2pt j}$. Here $i,j$ are
the matrix labels of the gauge group generators
$(T^{\mathcal {A}})_i^{\hskip2pt j}$.
It follows that we can construct non-local conserved charges by
\begin{equation}
\label{nsc}
{Q^{AB...}_0 = \int_{M} \kappa^A_\nu
(\theta^{0\nu})_i^{\hskip2pt j} \int_{M} \kappa^B_\rho
(\theta^{0\rho})_j^{\hskip2pt k}\ldots\,,}
\end{equation}
where $M$ is an initial
value surface in spacetime. In free field theory, this acts on a
chain of partons rather as (\ref{nobbo}) does, but we have no idea how
to extend the definition to $g^2N\not= 0$.

Inspired by the above examples,
our basic assumption in [\refcite{DNW}] is that in ${\mathcal N} =4$ super
Yang-Mills theory at $g^2N=0$, with $J_i^A$ understood as the
$PSU(2,2|4)$ generators of the $i^{th}$ parton, (\ref{nobbo}) is the
correct formula for the Yangian generators $Q^A$. Our assumption,
in other words, is that the bilocal symmetry deduced from [\refcite{Bena}]
goes over to (\ref{nobbo}) for $g^2N\to 0$. Of course, in any case
(\ref{uncu}) is the appropriate free field formula for the $J^A$, so we
do not need to state any hypothesis for these generators. And no
further assumption is needed for the higher charges in the
Yangian; they are generated by repeated commutators of the $Q^A$.
So our hypothesis about $Q^A$ completely determines the form of
the Yangian generators in the free-field limit.

Though our goal in these notes is to fill the above-mentioned
technical gap in the previous analysis, for completeness we here
sketch some of the reasoning in our previous paper. (We return in
the next section to the question of why the simple bilocal formula
does give a representation of the Yangian.) Having made an ansatz
for how the Yangian algebra is realized at $g^2N=0$, we consider
what happens when $g^2N$ is not quite zero. Some generators of the
Yangian do not receive quantum corrections. For example, the
spatial translation symmetries and the Lorentz generators are
uncorrected, because the theory can be regularized in a way that
preserves them. But the dilatation operator $D$ -- the generator
of scale transformations -- certainly is corrected. The
corrections to the eigenvalues of $D$ are called anomalous
dimensions.

We assume, in view of [\refcite{Bena}], that the ${\mathcal N}=4$ Yang-Mills
theory in the planar limit does have Yangian symmetry for all
$g^2N$. If so, the corrections modify the form of the generators,
but preserve the commutation relations.  One of the commutation
relations says that $Q^A$ transforms in the adjoint representation
of the global group $PSU(2,2|4)$ generated by $J^A$:
$[J^A,Q^B]=f^{ABC}Q^C$. We will write $J^A$ and $Q^A$ for the
charges at $g^2N=0$, and $\delta J^A$ and $\delta Q^A$ for the
corrections to them of order $g^2N$. We write ${\tilde J}^A$ and
${\tilde Q}^A$ for the exact generators (which depend on $g^2N$), so
${\tilde J}^A=J^A+(g^2N)\delta J^A+{\mathcal{O}}((g^2N)^2)$, and likewise
for ${\tilde Q}^A$. To preserve the commutation relations, we have

\begin{equation}
\label{incoco}{[\delta J^A,Q^B]+[J^A,\delta Q^B]=f^{ABC}\delta Q^C.}
\end{equation}
We are now going to make an argument for the Yangian that
parallels one used in [\refcite{Beisertone}] for the $PSU(2,2|4)$
generators. We consider the special case of this relation in which
$A$ is chosen so that $J^A$ is the dilatation operator $D$. We
also pick a basis  $Q^B$ of the $Q$'s to diagonalize the action of
$D$, so the $PSU(2,2|4)$ algebra reads in part
$[D,Q^B]=\lambda^BQ^B$, where $\lambda^B$ is the bare conformal
dimension of $Q^B$.  Then (\ref{incoco}) gives us
\begin{equation}
\label{pico}
{[\delta D,Q^B]+[D,\delta Q^B] = \lambda^B \delta Q^B.}
\end{equation}
However, in
perturbation theory, operators only mix with other operators of
the same classical dimension.  So just as $[D,Q^B]=\lambda^B Q^B$,
we have $[D,\delta Q^B]=\lambda^B \delta Q^B$. Combining this with
(\ref{incoco}), we have therefore
\begin{equation}
\label{tuggo}
{[\delta D,Q^B]=0.}
\end{equation}
Precisely
the same argument was used in [\refcite{Beisertone}] to show that $[\delta
D,J^A]=0$; this was a step in determining $\delta D$. Combining
this with (\ref{tuggo}), we see that $\delta D$ must commute with the
$g^2N=0$ limit of the whole Yangian.

The structure of perturbation theory implies in addition that the
operator $\delta D$ is a sum of operators local along the chain;
this fact has been exploited in [\refcite{bmn}] and many subsequent papers.
(In fact, $\delta D$, as described explicitly in [\refcite{Beisertone}], is a
sum of operators that act on nearest neighbor pairs.)  The
operators of this type that commute with the Yangian -- where here
we mean the Yangian representation most commonly studied in
lattice integrable models, which for us is the one generated at
$g^2N=0$ by $J^A$ and $Q^A$ -- are called the Hamiltonians of the
integrable spin chain. Thus, from our assumption about the
free-field limit of the Yangian, we are able,  starting with
the nonlocal symmetries found in [\refcite{Bena}], to deduce the basic
conclusion of [\refcite{Beiserttwo}], found earlier in a special case in
[\refcite{Minahan}],  that $\delta D$ is a Hamiltonian of an integrable spin
chain.

In our paper [\refcite{DNW}], we verify this picture by
proving directly, using formulas for the one loop operator computed in
[\refcite{Beisertone}], [\refcite{Beiserttwo}], that it is true that $\delta D$
commutes with the Yangian.  Since its commutativity with $J^A$ was
already used in [\refcite{Beisertone}] to compute $\delta D$, the only novelty is 
to verify that $[\delta D, Q^A]=0$.

{}From what we have said, it is clear that the appearance of a
Hamiltonian that commutes with the Yangian depends on expanding to
first order near $g^2N=0$.  In the exact theory, at a nonzero
value of $g^2N$, one would simply say that the exact dilatation
operator ${\mathcal D}$, which of course depends on $g^2N$, is one
of the generators of the Yangian. It is not the case in the exact
theory that one has a Yangian algebra and also a dilatation
operator that commutes with it.

This result is highly non trivial, and is heavily based on non-trivial
properties of the loop correction to the dilaton operator.
Hence, it is a good step in the direction of proving that (\ref{nobbo})
is the correct Yangian charge. To strengthen our guess for the $Q^A$ charges
as expressed in (\ref{nobbo}),
we will show that they satisfy the Serre relation, or equivalently
the nesting relation given in (\ref{yangs}).
We first explain the key steps for $SU(N)$, and then outline how
these steps change and generalize
in the case of $PSU(2,2|4)$ under consideration.

\section{Yangian Relations for $SU(N)$}

We return to the question of showing that the bilinear ansatz
(\ref{nobbo}) does give, under certain conditions, a solution of the
Serre relations. We will show explicitly that (as indicated in
[\refcite{Bernardtwo}]) the standard relations for the Yangian
$Y(G)$, which are valid for any Lie group G, are equivalent when
$G$ is $SU(N)$ (or $U(N)$) to a matrix form of the commutation
relations. We then use this to show that for certain types of
representation ${\mathcal R}$, the formula (\ref{nobbo}) does give a solution
of the Serre relations.

For $G = SU(N)$, the Lie algebra is the space of traceless
$N\times N$ matrices.  Instead of describing the Lie algebra in
terms of an abstract basis $J^A$, as one could do for any Lie
group $G$, it is useful for $SU(N)$ to describe it in terms of
generators $J^a{}_b$, $a,b=1,\dots, N$, with $\sum_aJ^a{}_a=0$,
and obeying
\begin{eqnarray}
&&[J^a{}_b, J^c{}_{d}] = \delta^c{}_b J^a{}_d - \delta^a{}_d J^c{}_b
\,,\nonumber\\
&&[J^a{}_b, Q^c{}_d] = \delta^c{}_b \,Q^a{}_d -
\delta^a{}_d \,Q^c{}_b\,. \label{comrel}
\end{eqnarray}
Similarly, the generators $Q^A$ of the Yangian are rewritten as a
traceless $N\times N$ matrix of operators $Q^a_b$. The Serre
relation then becomes
\begin{eqnarray}
&& [J^a{}_b, [ Q^c{}_d, Q^e{}_f]] - [Q^a{}_b, [J^c{}_d, Q^e{}_f]]
\nonumber\\
&& =  {h^2\over 4} \sum_{p,q} (\, [ J^a{}_b, [J^c{}_p
J^p{}_d, J^e{}_q J^q{}_f]] - [J^a{}_p J^p{}_b,
[J^c{}_d, J^e{}_q J^q{}_f]]\,)\,. \label{span}
\end{eqnarray}
This matrix form of the Yangian relations is the most familiar one
in integrable systems, and will be useful in the generalization to
the superalgebra. We will prove below that (\ref{span}) is
equivalent to (\ref{yangs}).

To do this, it is useful to work out in more detail how general Lie algebra
notation simplifies for
$SU(N)$.  The generators $T^A$, $A=1,\dots,N^2-1$, of $SU(N)$  can be regarded
as $N\times N$ matrices in the fundamental representation of
$SU(N)$.  We use the conventions $[T^A, T^B] = f^{AB}_C T^C$,
$f_{ABC} f_{ABE} = 2 N \delta_{CE}$,
\begin{eqnarray}
\Tr T^A T^B &&= - \delta^{AB} \nonumber \\
\Tr  T^A T^B T^C &&= -\half (f_{ABC} -i d_{ABC})\,.
\label{tcs}
\end{eqnarray}

We define the totally symmetric invariant tensor $i d_{ABC} = Tr
(\{T^A, T^B\} T^C)$, and use the fact that the generators in this
representation, together with the identity matrix, span the space
of all complex $N \times N$ matrices.  We get:
\begin{eqnarray}
T^A T^B &&= \half (f_{ABC} T^C + \{T^A, T^B\}) =
-{1\over N} \delta_{AB} + \half (f_{ABC} -i d_{ABC}) T^C\nonumber \\
{}[T^A, T^B] &&= f_{ABC} T^C\nonumber\\
\{T^A, T^B\} &&= -{2\over N}  \delta_{AB} -i d_{ABC} T^C\,.
\label{tcsb}
\end{eqnarray}
Explicitly, we write the matrix elements of the matrix $T^A$, in
the fundamental representation, as $T^A{}^a{}_b$, $a,b=1\dots N$.
We define
\begin{equation} J^A = -T^A{}^{b}_{a} J^{a}_{b}\,,\qquad Q^A = -T^A{}^{b}_{a}
Q^{a}_{b}\,, \label{sm}
\end{equation}
where the two-index generators in (\ref{sm}) are traceless $\sum_a
J^{a}_{a} = 0 = \sum Q^{a}_{a}$, and we can invert $J^{a}_{b} =
J^A T^A{}^{a}_{b}$ and  $Q^{a}_{b} = Q^A T^A{}^{a}_{b}$. Note that
although $T^A{}^{a}_{b}$ is in the $N$ of $SU(N)$, $J^A$ and
consequently $J^{a}_{b}$ can be in an arbitrary representation.
The  relation (\ref{span}), which is sometimes called the nesting
relation, reduces to (\ref{yangs}) as follows. Multiplying
(\ref{span}) by $ T^A{}^{b}_{a} T^B{}^{d}_{c} T^C{}^{f}_{e}$, we
find
\begin{eqnarray}
&&[Q^A, [Q^B, J^C]]
+ [Q^B, [Q^C, J^A]] + [Q^C, [Q^A, J^B]]\nonumber\\
&&=  - {h^2\over 4} \,(\, (\Tr T^B T^D T^E) \, \Tr (T^C T^F T^G
)\,
[J^A, [ J^D J^E, J^F J^G]]]\nonumber\\
&&\hskip20pt - (\Tr T^A T^D T^E) \, \Tr (T^C T^F T^G )\,
[J^D J^E, [ J^B, J^F J^G]]\,)]\nonumber\\
&&=  {h^2\over 16} \,(\, - f_{ACG} f_{GLM} ( d_{MDE} d_{BLF}
+ d_{MFE} d_{BLD} )\nonumber\\
&&\hskip35pt + f_{ABG} f_{GLM} ( d_{MDE} d_{CLF} + d_{MFE} d_{CLD})\nonumber\\
&&\hskip35pt + f_{BCG} f_{GLM} ( d_{MDE} d_{ALF} + d_{MFE} d_{ALD}))
J^D J^E J^F\,.
\label{spanone}
\end{eqnarray}
To evaluate the products of $d$ symbols and structure constants
solely in terms of the structure constants as in (\ref{yangs}), we
use the following identities. In addition to the Jacobi identity,
there is a similar formula $[\{T^A,T^B\}, T^C] + [\{T^C,T^A\},
T^B] + [\{T^B,T^C\}, T^A] = 0,$ which reduces to
\begin{equation}
d_{ABE} f_{ECD} + d_{CAE} f_{EBD} + d_{BCE} f_{EAD} = 0\,.
\label{jacd}
\end{equation}
Another identity
$[[T^A,T^B], T^C] + \{\{T^C,T^A\}, T^B\} -  \{\{T^B,T^C\}, T^A\} = 0$
results in
\begin{equation}
f_{ABE} f_{CDE} = {4\over N} (\delta_{AC} \delta_{BD}
-\delta_{BC} \delta_{AD} )
+ d_{ACE} d_{BDE} - d_{BCE} d_{ADE}\,,
\label{tdtf}
\end{equation}
and
\begin{equation}
d_{ABC} d_{ABE} = 2 ( N - {4\over N})\, \delta_{CE}\,.
\label{tdal}
\end{equation}
{}From the Jacobi identity, we find the triple product
\begin{equation}
f_{DMA} f_{ABE} f_{ECD} = - N f_{MBC}\,,
\label{trpdone}
\end{equation}
from (\ref{jacd}) we find
\begin{equation}
d_{ABE} f_{ECD} f_{DFA} = - N d_{BCF}\,
\label{trpdt}
\end{equation}
and from (\ref{tdtf}), and $\sum_A d_{AAB} = 0$, we have
\begin{eqnarray}
d_{ACE} d_{EBD} f_{DMA} &&= (N-{4\over N}) f_{CBM}\nonumber\\
d_{DMA} d_{ACE} d_{EBD} &&= ( N - {12\over N}) d_{MCB}\,.
\label{trpdthr}
\end{eqnarray}
Note that (\ref{tdtf}) expresses the difference of two products of
$d$ symbols in terms of the structure constants.
We used  (\ref{jacd}) in deriving (\ref{spanone}). It is
convenient to symmetrize on the $D,E,F$ indices in (\ref{spanone})
and use (\ref{trpdt},\ref{trpdthr}). Then (\ref{spanone}) becomes
\begin{eqnarray}
&&[Q^A, [Q^B, J^C]]
+ [Q^B, [Q^C, J^A]] + [Q^C, [Q^A, J^B]]\nonumber\\
&&= {h^2\over 48} \,(\,f_{ABG} f_{GLM} d_{MDE} d_{CLF}
-(A\leftrightarrow C)
-(B\leftrightarrow C)) \{J^D, J^E, J^F\}\,,\qquad\ast\nonumber\\
&&=  {h^2\over 48} \,(\,(f_{ALG} f_{GBM} + f_{BLG} f_{GAM})  d_{MDE} d_{CLF}
-(A\leftrightarrow C)
-(B\leftrightarrow C))\cr
&&\hskip35pt  \{J^D, J^E, J^F\}\nonumber\\
&&=  {h^2\over 24} \,(\,(( f_{FAL} f_{DBM}  d_{CLG} d_{GME}
-(A\leftrightarrow B))  -(A\leftrightarrow C) -(B\leftrightarrow C))\nonumber\\
&&\hskip35pt
+  2 (f_{CAL} f_{DBM} d_{LFG} d_{GME}
-(A\leftrightarrow B) -(B\leftrightarrow C))) \,\{J^D, J^E, J^F\}\nonumber\\
&&=  {h^2\over 24} \,(\,(( f_{FAL} f_{DBM}  d_{CLG} d_{GME}
-(A\leftrightarrow B))  -(A\leftrightarrow C) -(B\leftrightarrow C))\nonumber\\
&&\hskip35pt
- (f_{BAL} f_{GLM} d_{CFG} d_{MDE}
-(A\leftrightarrow C) -(B\leftrightarrow C))) \,\{J^D, J^E, J^F\}\,\ast
\nonumber\\
\label{spantwo}
\end{eqnarray}
where
$\{J^D, J^E, J^F\}$ is the totally symmetrized product,
\begin{eqnarray}
\{J^D, J^E, J^F\}&& =
J^D J^E J^F + J^E J^D J^F + J^F J^E J^D\cr
&&\hskip15pt  + J^E J^F J^D
+ J^D J^F J^E + J^F J^D J^E\,.
\label{symprod}
\end{eqnarray}
We observe that the two starred lines are proportional, so we have
\begin{eqnarray}
&& 3\, {h^2\over 48} \, (\,f_{ABG} f_{GLM} d_{MDE} d_{CLF}
-(A\leftrightarrow C)
- (B\leftrightarrow C)) \{J^D, J^E, J^F\}\qquad\noblackbox\nonumber\\
&&=  {h^2\over 24} \,(\,( f_{FAL} f_{DBM}  d_{CLG} d_{GME}
-(A\leftrightarrow B))  -(A\leftrightarrow C) -(B\leftrightarrow C))\cr
&&\hskip30pt \,  \{J^D, J^E, J^F\}\nonumber\\
&&=  {h^2\over 8}
\, f^{ADK}f^{BEL}f^{CFM}f_{KLM} \, \{J^D, J^E, J^F\}\,.
\label{spanthree}
\end{eqnarray}
It follows that
\begin{eqnarray}
&&[Q^A, [Q^B, J^C]]
+ [Q^B, [Q^C, J^A]] + [Q^C, [Q^A, J^B]]\noblackbox\nonumber\\
&&=  {h^2\over 24} \,  f^{ADK}f^{BEL}f^{CFM}f_{KLM} \, \{J_D, J_E,
J_F\}. \label{trion}
\end{eqnarray}
This equation  is totally antisymmetric in $A,B,C$, and for $h=1$
is the equation (\ref{yangs}).

\bigskip\noindent{\it A Useful Criterion}

The point of this lengthy analysis is that although it is
difficult to find a solution of the Yangian relation
(\ref{yangs}), it is much easier to find a solution of
(\ref{span}).  The basic case is the case of just a single spin.
We want to find a criterion under which, for $G=SU(N)$ and some
irreducible representation ${\mathcal R}$ at a single site, we can obey the
Yangian algebra with the simple choice $Q^A=Q^a{}_b=0$.  In
(\ref{yangs}), it is unclear when this works, but in (\ref{span}),
we can easily find a simple criterion.

Since the left hand side of (\ref{span}) obviously vanishes for
$Q^A=0$, we need a criterion for vanishing of the right hand side.
The object $J^c{}_pJ^p{}_d$ that appears in (\ref{span}) is a
linear combination of pieces that transform as the singlet of
$SU(N)$ and the adjoint.  If ${\mathcal R}$ is irreducible, the singlet piece
is a multiple of the identity and does not contribute in the
commutator.  If moreover ${\mathcal R}$ is such that the adjoint only appears
once in the decomposition of ${\mathcal R}\otimes \overline {\mathcal R}$, 
then (modulo the irrelevant $c$-number) $J^c{}_pJ^p{}_d$ is a multiple of
$J^c{}_d$.  Similarly, $J^e{}_qJ^q{}_f$ can be replaced by the
same multiple of $J^e{}_f$, and $J^a{}_pJ^p{}_b$ by the same
multiple of $J^a{}_b$.  Once this is done, the right hand side of
(\ref{span}) vanishes. 

So we have a criterion for finding irreducible representations ${\mathcal R}$
of $SU(N)$ such that for a single-spin system, the Yangian algebra
is obeyed with $Q^A=Q^a{}_b=0$.  This criterion is not obeyed for
all representations.  For example, if ${\mathcal R}$ is the adjoint
representation, then the criterion is not obeyed, since the
adjoint then appears twice in the decomposition of ${\mathcal R}\otimes
\overline{\mathcal R}$ 
(which in this example is the same as ${\mathcal R}\otimes {\mathcal R}$).

However, many important examples arise from representations ${\mathcal R}$
that {\it do} obey the criterion.  Basic examples are the
fundamental representation, and more generally the representation
of antisymmetric $k^{th}$ rank tensors, for any $k$.

\bigskip\noindent{\it A Chain Of Spins}

Now let us explain how to go from this single-spin result to a representation
of the Yangian algebra on a whole chain of spins.

The important property of the Yangian algebra is that it is a Hopf
algebra, which means that there is a natural recipe for defining a
tensor product of representations.  If ${\mathcal A}$ is an algebra, a
``coproduct'' is a map $\Delta:{\mathcal A}\to {\mathcal A}\otimes {\mathcal
A}$ that is a homomorphism of algebras (and obeys some additional axioms of which
we explain the relevant one later).  For our purposes, this
means that the coproduct maps  operators that represent ${\mathcal A}$
in a single-spin representation ${\mathcal R}$ to operators that represent 
${\mathcal A}$ in
the Hilbert space ${\mathcal R}\otimes {\mathcal R}$ of a two-spin system.  
Moreover,
one can repeat the process, using the homomorphism $\Delta\otimes
1:{\mathcal A}\otimes {\mathcal A}\to {\mathcal A}\otimes {\mathcal A}\otimes
{\mathcal A}$, or alternatively the homomorphism $1\otimes \Delta:
{\mathcal A}\otimes {\mathcal A}\to {\mathcal A}\otimes 
{\mathcal A}\otimes {\mathcal A}$,
to map a representation in the two-spin system to a
representation in the three-spin system. (Here, for example, $1\otimes \Delta$
is the operator that acts as the identity on the first spin and acts by 
$\Delta$
to map the Hilbert space of the second spin in a two-spin system to that of the
last two spins in a three-spin system.  Similarly $\Delta\otimes 1$ acts as the
identity on the second or last spin while mapping the Hilbert space of the first
to that of a two-spin system.)   Repeating the process,
one gets a representation of ${\mathcal A}$ in an $n$-spin system, for any $n$.
Apart from being a homomorphism of algebras, the key axiom obeyed
by $\Delta$ is ``coassociativity,'' $\Delta\otimes
1(\Delta)=1\otimes \Delta(\Delta)$, which for our purposes says
that starting with a given representation of ${\mathcal A}$ in the
single-spin system, the representation that one arrives at in the
$n$-spin system does not depend on the the precise route by which
one  applies these formulas.

For the Yangian, the explicit formula for the coproduct is
\begin{eqnarray}
\Delta(J^A)&&=J^A\otimes 1+1\otimes J^A\,\cr
\Delta(Q^A)&&=Q^A\otimes 1+1\otimes Q^A+f^A_{BC}J^B\otimes J^C\,.
\label{polo} 
\end{eqnarray}
Using this coproduct, we can determine, given a single-spin
representation of the Yangian with $Q^A=0$, what the
representation should be for a multi-spin system.  Consider first
the two-spin system.  The two-spin representation of $J^A$ is
$\Delta(J^A)=J^A\otimes 1+1\otimes J^A$.  This is a fancy notation for writing
the result that we would expect naively, since
$J^A\otimes 1$ and $1\otimes J^A$ are simply in a more typical physics notation
the operators
$J_1^A$ and $J_2^A$ that
act by $J^A$ on the first or second spin. So $\Delta(J^A)=J^A_1+J^A_2$, saying
simply that the group generators of the two-spin system are the sum of the 
single-particle generators. If the single-spin
representation of $Q^A$ is zero, then the two-spin
representation of $Q^A$ reduces to
$\Delta(Q^A)=f^A_{BC}J^B\otimes J^C$, which in the alternative notation
is $f^A_{BC}J_1^BJ_2^C$. This agrees with the two-spin case of
(\ref{nobbo}).  More generally, by repeated application of the coproduct, one
learns that whenever for the one-spin system one can obey the
Yangian algebra with $Q^A=0$, the formula (\ref{nobbo}) supplies
a representation of the Yangian algebra for a chain of spins.

If the adjoint representation appears more than once in the
decomposition of ${\mathcal R}\otimes \overline {\mathcal R}$, 
then for the one-spin
system, one cannot generally obey the Yangian algebra with
$Q^A=0$. However, for $SU(N)$, the form (\ref{span}) of the
relations implies that one can always, for any representation ${\mathcal R}$ 
of $SU(N)$, obey the Yangian algebra
with $Q^a{}_b=J^a{}_pJ^p{}_b/2$.  This
contrasts with the situation for more general symmetry groups $G$,
where for generic ${\mathcal R}$ there is no choice of $Q^A$ that obeys the
Serre relation.

\section{Yangian Superalgebra}

The analysis in the previous section would have worked in just the
same way if we replace the simple group $SU(N)$ by  the non-simple
group $U(N)$.  It similarly works for the supergroup $U(N|M)$, and
for $SU(N|M)$ if $N\not= M$.  The case $N=M$, however, requires a
further study.

This fact is relevant for us because $PSU(2,2|4)$, which is a real
form of $PSU(4|4)$, is the symmetry of ${\mathcal N}=4$ super
Yang-Mills theory, which thus involves the exceptional case
$N=M=4$. (For our purposes, the signature is not important, as we
will be carrying out purely algebraic manipulations; we need not
distinguish $PSU(4|4)$ from $PSU(2,2|4)$.)

First we give a few facts about the Lie superalgebras $U(n|n),
SU(n|n)$ and $ PSU(n|n)$ (see {\it eg.} [\refcite{BVW}]). 
For references to super Yangians, see {\it eg.} [\refcite{Nazarov}-
\refcite{Witten}]. The Lie
superalgebra $U(n|m)$ has generators that can be represented
by matrices of the form $ x = \left(\begin{array}{cc}a&b\\
c&d \end{array}\right)$, where $a$ and $d$ are $n\times n$ and
$m\times m$ bosonic hermitian matrices, and $b$ and $c$ are
$n\times m$ and $m\times n$ fermionic matrices and are hermitian
conjugates. The supertrace of $x$ is Str $x$ $=$ Tr $a$ $ -$ Tr $d$.
If we divide by multiples of the identity, we get a superalgebra
$PU(n|m)$. If we restrict to $x$ such that Str $x$$=0$, we get a
superalgebra of one less dimension that is called $SU(n|n)$.
For $n\not= m$, at the Lie algebra level, requiring the trace of $x$
to be zero removes the identity matrix and hence 
$PU(n|m)$ and $SU(n|m)$ are the same at the Lie algebra level
(the global structure of the groups is different).  For $n=m$, the
identity matrix has zero supertrace, so requiring the trace to be
zero does not remove the identity matrix.  If we require $x$ to be
traceless and further identify any two $x's$ that differ by an
additive scalar, we get a  superalgebra that is called $PSU(n|n)$
or $ A(n-1|n-1)$ and has two dimensions less than $U(n|n)$.

The generators of  $PSU(n|n)$ can be represented by matrices
$x =  \left(\begin{array}{cc}a&b\\
c&d \end{array}\right)$, with Tr $a$ $=$ Tr $d$ $=0$. The bosonic
part of $PSU(n|n)$ is $SU(n)\times SU(n)$, generated by $a$ and
$d$. We are here writing  $x$  in what we may call the $n|n$
representation. In any representation, the fermionic generators of
$PSU(n|n)$, here represented by the matrices $b$ and $c$,
transform as $n\otimes\bar n \oplus \bar n\otimes n$ under
$SU(n)\times SU(n)$.

The superconformal symmetry group of ${\mathcal N}=4$ super Yang-Mills
theory is  a real form of $PSU(4|4)$, whose bosonic part is
$SU(4)\times SU(4)$. 
\def\Str{{\rm Str}}
It will be helpful to compare $PSU(4|4)$ to its cousins $SU(4|4)$
and $U(4|4)$. We have for the Lie algebras
\begin{eqnarray}
SU(4|4) &&= PSU(4|4) \oplus R\cr U(4|4) &&=  PSU(4|4) \oplus K
\oplus R\,. \label{salg}
\end{eqnarray}  (This is an additive decomposition of the
Lie algebras; the commutation relations do not preserve this
decomposition.) $K$ is the Lie algebra generated by
the identity matrix (which we also write as $K$). $R$ 
is the Lie algebra of a $U(1)$ $R$-symmmetry group 
that is not contained in $PSU(4|4)$ and is not a symmetry of
${\mathcal N}=4$ super Yang-Mills theory; 
we also call its generator $R$.  In the $4|4$
representation, we take $Q_R = R$ where
\begin{equation} R={1\over 2}\left(\begin{array}{cc}1&0\\0&-1\end{array}
\right)\,,
\label{balg}
\end{equation} 
So commutation with $R$
multiplies the blocks $b$ and $c$ of a  generator $x$ by $1$ or
$-1$ and annihilates $a$ and $d$. The supertraces are
\begin{equation} \Str\,K^2=\Str\,R^2=0,
~\Str\,RK=4.\label{calg}\end{equation}

We define the $U(4|4)$ structure constants by
\begin{equation}
[J_A, J_B\} = f^C_{AB} J_C\,, \label{scomrel}
\end{equation}
where the brackets denote either commutators or anticommutators.
Then $f^A_{KB}=0$ for all $A,B$ since $K$ is central and commutes
with everything, and $f^R_{AB}=0$ for all $B,C$, since the $U(1)$
$R$-symmetry generator $R$ never appears on the right hand side of
the commutation relations. (It is precisely because $R$ never
appears on the right hand side of the commutation relations that
there can exist a theory, such as ${\mathcal N}=4$ super Yang-Mills
theory, that has $PSU(4|4)$ symmetry but not the additional $U(1)$
$R$-symmetry.)

The formula (\ref{scomrel}) is the first formula in this paper in which
it is important to carefully distinguish whether the ``$A$'' 
index of a Lie algebra
generator such as $J_A$ is ``down'' or ``up.''  At the outset of
these notes, we merely asserted that there is an invariant,
nondegenerate metric $g$  that is used to raise and lower indices, and in many
formulas we have done so without comment.
In the present example, we can take the metric for $U(4|4)$ to be
$g_{AB}=\half \Str \,J_AJ_B.$  So $g_{KK}=g_{RR}=0$,
$g_{KR}\not=0$; 
and when $A\ne K,R$ then $g_{AB} = 2 \delta_{AB}$ and $g_{KA} = g_{RA} = 0$.
It follows that when we raise and lower
indices, $K$ and $R$ are exchanged, so the assertions in the last
paragraph become
\begin{equation}   f_A{}^{RB}=0=
f_K^{AB}.\label{dumbo}\end{equation}

For $U(4|4)$, the analysis in the last section applies and shows
that the simple bilinear formula (\ref{nobbo})
 gives a representation of the Yangian algebra as
long as the single-spin representation ${\mathcal R}$ has the property that
the adjoint representation only appears once in the decomposition
of ${\mathcal R}\otimes \overline {\mathcal R}$.  
For ${\mathcal N}=4$ super Yang-Mills
theory, we take ${\mathcal R}$ to be the representation consisting of the
one-particle states of the
free vector multiplet.  The $U(1)$ $R$-symmetry generator $R$ does
act on this representation (though it is not a symmetry of the
gauge theory), and we take $K$ to act on the representation by
$K=0$.  In this way we interpret the representation ${\mathcal R}$ as a
representation of the extended group $U(4|4)$.

This representation does have the property that the adjoint
representation only appears once in the decomposition of ${\mathcal R}\otimes
\overline {\mathcal R}$.  So we can use the familiar bilinear formula to get
a multi-spin representation of the Yangian of $U(4|4)$:
\begin{eqnarray}
Q_C &&= g_{CC'}Q^{C'}=g_{CC'}f^{C'}_{AB} \sum_{i<j} J_i^A J_j^B\nonumber\\
&&= g_{CC'}f^{C'}_{AB} g^{AA'} g^{BB'}  \sum_{i<j} (J_i)_{A'}
(J_j)_{B'}. \label{nlgen}
\end{eqnarray}
Here we have been careful in raising and lowering of indices to
ensure that the second generator $Q_C$ of the Yangian transforms
like the $PSU(4|4)$ generators $J_C$.

Since we actually want to represent the Yangian of $PSU(4|4)$, 
not that of $U(4|4)$,
we need a few more observations.
From (\ref{nlgen}), we see  that in the representation of the $PSU(4|4)$
Yangian, $Q_K = 0$ and $Q_R = 2 f^K_{AB}
g^{AA'} g^{BB'} \sum_{i<j} (J_i)_{A'} (J_j)_{B'}$. What about the
generators $Q_C$ where $C$ corresponds to a generator of
$PSU(4|4)$?  They do not depend on $K$, since $K=0$ in our chosen
representation, and they do not depend on $R$, since
$f^{C'}_{KB}=0$.  It follows that the $Q_C$'s are given by the
same formula as if we had evaluated the bilinear formula (\ref{nobbo})
directly for $PSU(4|4)$. 
We have thus established that this bilinear formula does give a 
representation of the Yangian algebra for $PSU(4|4)$.

\end{document}